\documentclass{webofc}
\usepackage[varg]{txfonts}   

\usepackage{amsmath}
\usepackage{amssymb}
\usepackage{graphicx}
\usepackage{subfigure}
\usepackage{babel}
\usepackage{enumerate}

\usepackage{tikz}

\newcommand{\Mev}{MeV/A }
\newcommand{\mev}{MeV/A}

\newcommand{\Xesn}{$^{129}$\text{Xe}+$^{nat}$\text{Sn} }

\newcommand{\Dt}{$\delta t$ }

\newcommand{\imgpath}{.}
\renewcommand{\bf}[1]{\begingroup\bfseries\mathversion{bold}#1\endgroup}

\begin{document}

\woctitle{Fifth International Workshop on Nuclear fission and Fission-Product Spectroscopy}

\title{Sequential fissions of heavy nuclear systems}

\author{
        D. Gruyer\inst{1}\fnsep\thanks{\email{diego.gruyer@ganil.fr}}  \and J.D. Frankland\inst{1}  \and  E. Bonnet\inst{1} \and M. Boisjoli\inst{1,2} \and  A. Chbihi\inst{1}
        \and L. Manduci\inst{3}  \and P. Marini\inst{1} \and  K. Mazurek\inst{4} \and P.N. Nadtochy\inst{5}~ for the INDRA collaboration
}

\institute{
           GANIL, CEA-DSM/CNRS-IN2P3, Bvd. Henri Becquerel, F-14076 Caen Cedex, France 
\and 
D\'epartement de physique, de g\'enie physique et d'optique, Universit\'e Laval, Qu\'ebec, G1V 0A6 Canada
\and 
           \'Ecole des Applications Militaires de l'\'energie Atomique, B.P. 19, F-50115 Cherbourg, France
\and 
           The Niewodnicza\'nski Institute of Nuclear Physics - PAN, PL-31-342 Krak\'ow, Poland
\and 
           Omsk State University, Mira prospekt 55-A, Omsk 644077, Russia
          }

\abstract{%
In \Xesn central collisions from 12 to 20 \Mev measured with the INDRA 4$\pi$ multidetector, the three-fragment
exit channel occurs with a significant cross section.
In this contribution, we show that these fragments arise from two successive binary splittings of a heavy composite system. 
Strong Coulomb proximity effects are observed
in the three-fragment final state. By comparison with Coulomb trajectory calculations,
we show that the time scale between the consecutive break-ups
decreases with increasing bombarding energy, becoming compatible with quasi-simultaneous multifragmentation above 18 \mev.
}

\maketitle

\section{Introduction}

In central heavy-ion collisions at beam energies between 25 and 100 \mev,  
production of many nuclear fragments is observed. The fragment production is compatible with the
simultaneous break-up of finite pieces of excited nuclear matter \cite{Borderie2008551}. 
This so-called ``nuclear multifragmentation'' is a fascinating process which has been widely studied by the INDRA collaboration,
notably in \Xesn central collisions 
\cite{Marie199715,Rivet1998217,PhysRevLett.86.3252,PhysRevC.67.064613,PhysRevC.71.034607,Tabacaru2006371,LeNeindre200747,Piantelli2008111,Bonnet20091,PhysRevC.86.044617}. 
But the energy required for the onset of multifragmentation is still an open question.

With the recent data on \Xesn reaction at energies between 8 and 20 \mev, 
Chbihi et al. \cite{Chbihi012099} have shown that at the lowest beam energy (8 \mev) central collisions lead mainly to two fragments in the exit channel.
From 12 \mev, the three-fragment exit channel becomes significant. However these fragments might be produced by sequential fission \cite{Bizard1992413}, ternary fission \cite{Herbach2002207} or 
multifragmentation \cite{Borderie2008551}. 
In this contribution, we determine the order and time scale of three fragment emission and show the evolution of the deexcitation process 
from hot sequential fission to multifragmentation.

\section{Experimental details}

Collisions of \Xesn at 12, 15, 18, 20 \Mev were measured using the INDRA $4\pi$ charged
product array \cite{Pouthas1995418} at the GANIL accelerator facility.
This detector, composed of 336 detection cells arranged 
in 17 rings centered on the beam axis, covers 90\% of the solid angle, and
can identify in charge fragments from hydrogen to uranium with low thresholds.

In this analysis, we considered only fusion-like events with 
three heavy fragments ($Z > 10$) in the exit channel.
The measured fragments in each event are sorted according to their atomic number such that $Z_1 \geqslant Z_2 \geqslant Z_3$.
This classification is introduced only to facilitate the presentation of the data.

\section{From sequential to simultaneous break-up}
\subsection{Qualitative evolution}
First we will show qualitatively the evolution of the decay process from two splittings well separated in time
towards simultaneous fragmentation.
If two successive independent splittings occur, three possible sequences of splittings have to be considered.
For instance, in one sequence, the first splitting leads to a fragment of charge $Z_1$ and another fragment which,
later, undergoes fission leading to $Z_2$ and $Z_3$.
Let us call this sequence \textit{1}. The sequences \textit{2} and \textit{3} are deduced by circular permutation.

\begin{centering}
\begin{figure}[ht]
     {\subfigure[ 12 MeV/A\label{fig:dlzP:12}]{\includegraphics[width=0.24\linewidth]{\imgpath/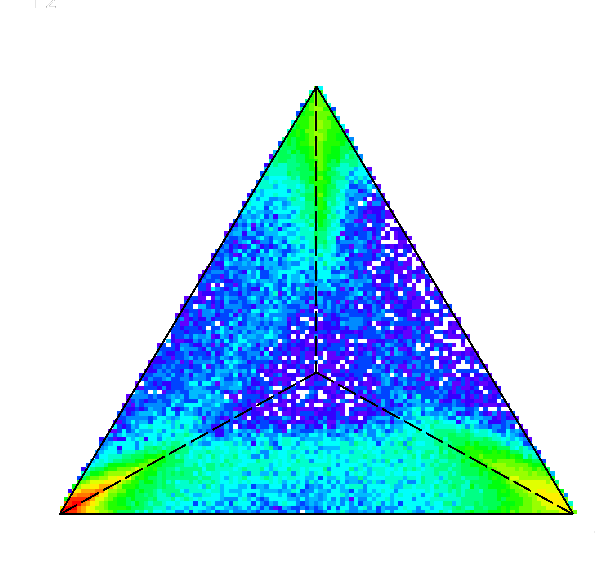}}}
     {\subfigure[ 15 MeV/A\label{fig:dlzP:15}]{\includegraphics[width=0.24\linewidth]{\imgpath/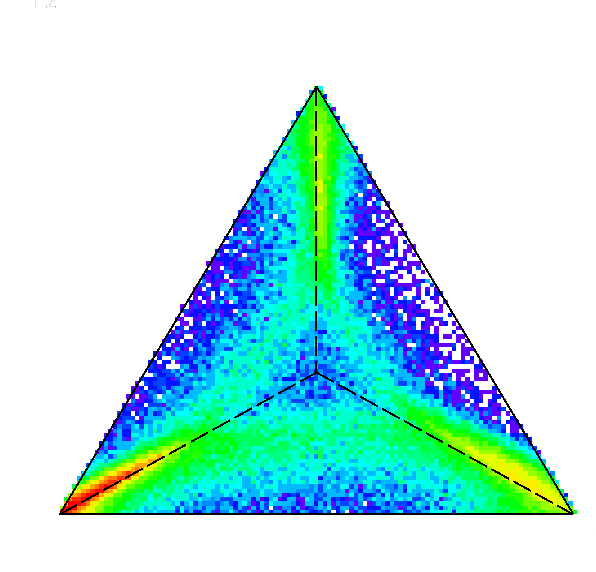}}}
     {\subfigure[ 18 MeV/A\label{fig:dlzP:18}]{\includegraphics[width=0.24\linewidth]{\imgpath/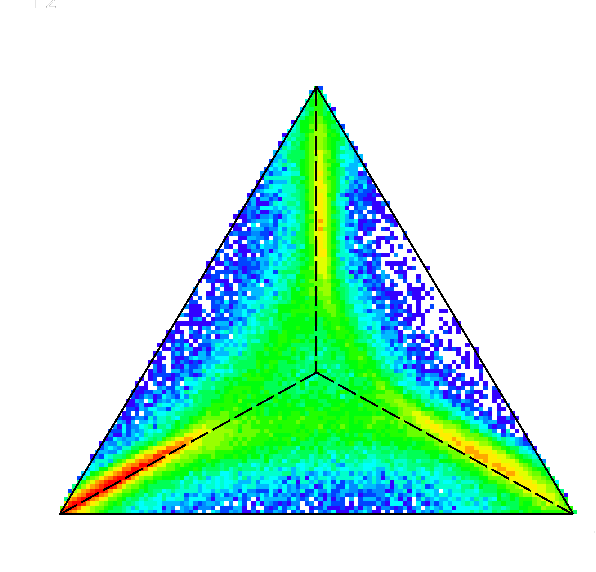}}}
     {\subfigure[ 20 MeV/A\label{fig:dlzP:20}]{\includegraphics[width=0.24\linewidth]{\imgpath/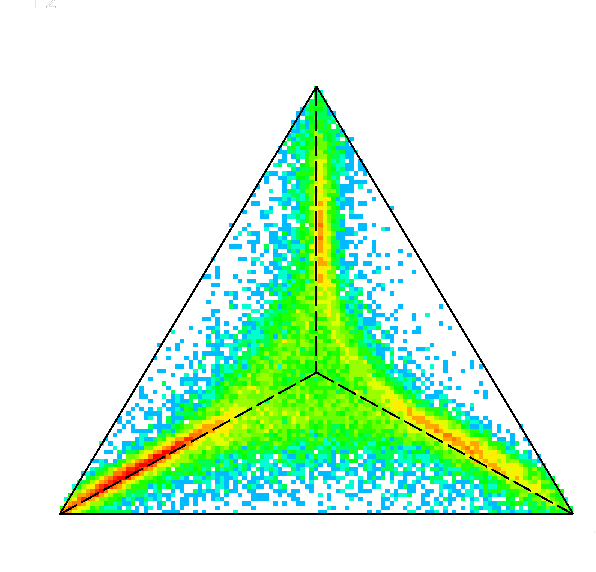}}}
\caption{(color online). Dalitz plot of $P_i$ (see text) for \Xesn central collisions at different beam energies.\label{fig:dlzP}}
\end{figure}
\end{centering}

Bizard et al. \cite{Bizard1992413} proposed a method to show qualitatively the nature of the process.
To test the compatibility of an event with the sequence of splittings \textit{i}, we compare the experimental
relative velocities with those expected for two successive fissions.
For each event we build the following quantities:
\begin{align}
 P_{i} = (v_{i(jk)}^{exp} - v_{i(jk)}^{viola})^2 + (v_{jk}^{exp} - v_{jk}^{viola})^2 
\label{eq:Bizard}
\end{align}
where $i$ = \textit{1}, \textit{2}, \textit{3}; $v_{\alpha\beta}^{exp}$ is the experimental relative velocity between fragments $\alpha$ and $\beta$; 
and $v_{\alpha\beta}^{viola}$ is the expected relative velocity for fission, taken from the Viola systematic \cite{PhysRevC.31.1550,Hinde1987318}. 
The first (second) term in Eq.(\ref{eq:Bizard}) refers to the first (second) splitting.
The lower the value of $P_i$, the larger the probability of the considered event to have been generated by the sequence of splittings
\textit{i}. The three values of $P_i$ are calculated for each event and represented in Dalitz plots (Fig.\ref{fig:dlzP}). 
In this diagram, the distance of each point from the three sides of the triangle reflects the relative values of $P_1$, $P_2$, and $P_3$.

\setcounter{subfigure}{0}

At 12 \Mev bombarding energy (Fig.\ref{fig:dlzP:12}), events populate mainly three branches parallel to the edges of 
the Dalitz plot, which correspond to the three sequences of sequential break-up ($P_i \ll P_j, P_k$). 
Simultaneous break-up events would be located close to the centre of this plot ($P_i \sim P_j \sim P_k$), where few events are observed.
The strong accumulations of events on the corners correspond to particular kinematic configurations where we are 
not able to disentangle two sequences ($P_i \sim P_j \ll P_k$).
Consequently, for this energy, fragments arise mainly from two sequential splittings.

With increasing beam energy (Fig.\ref{fig:dlzP}(b-d)), the three branches are still present but become closer and closer to the centre of the Dalitz plot. 
This means that fragment production becomes more and more simultaneous. In other words, when increasing the beam energy the deexcitation process 
evolves continuously from two sequential splittings towards simultaneous fragmentation.
In the following we will quantify this effect by measuring the time $\delta t$ between the two splittings.
First we must determine, event by event, in which order fragments have been produced.

 \subsection{Sequence of splittings}
To identify the sequence of splittings, we only consider the second separation step.
For each event, we compare the relative velocity between each pair of fragments with that expected for fission taken from the Viola systematics.
The pair with the most Viola-like relative velocity is considered to have been produced during the second splitting. 
We can therefore trivially deduce that the remaining fragment was emitted first.
This procedure amounts to computing, for each event, the three following values:
\begin{align}
 \chi_{i} = (v_{jk}^{exp} - v_{jk}^{viola})^2, \quad i=\textit{1, 2, 3,}
\label{eq:Moi}
\end{align}
which corresponds to the second term of Eq.(\ref{eq:Bizard}).
The smallest value of $\chi_{i}$ determines the sequence $i$ of splittings.

\begin{table}
\caption{Mean charges and charge asymmetries of the two splittings for \Xesn central collisions. E.C. refers to the entrance channel.\label{tab:charges}}
\center\begin{tabular}{r|c|c|c|c|c|c|c}
	 & $\langle Z_{src}\rangle$ & $\langle Z^f_{1}\rangle$ & $\langle Z^f_{2}\rangle$& \bf{$\langle \delta Z^{f}\rangle$}  & $\langle Z^s_{1}\rangle$ & $\langle Z^s_{2}\rangle$ & $\langle \delta Z^{s}\rangle$\\
\hline
 12 \mev & 88.8	& 25.5	& 63.3	& \bf{0.44}	& 40.0	& 23.3	& 0.26	\\
 15 \mev & 84.0	& 24.5	& 59.7	& \bf{0.43}	& 38.2	& 21.5	& 0.28	\\
 18 \mev & 79.9	& 24.0	& 55.8	& \bf{0.41}	& 35.8	& 20.0	& 0.28	\\
 20 \mev & 76.0	& 23.7	& 52.2	& \bf{0.40}	& 33.3	& 18.9	& 0.27	\\
 E.C.    & 104	& 50	& 54	& \bf{0.04}	& -	& -	& -	\\
\end{tabular}
\end{table}

\begin{figure}[ht!]
\sidecaption
\begin{tikzpicture}[scale=1]
\path [-,thin,opacity=1,color=black,dashed,above] (2+1,4) edge node {} (6.+1,4);
\path [-,thin,opacity=1,color=black,dashed,above] (4.5+1,3) edge node {} (5.5+1,5);
\path [->,thin,opacity=1,color=black,above] (2+1,4)   edge node {$\vec{v}^{f}_{1}$} (0.5+1,4);
\path [->,thin,opacity=1,color=black,right] (5.5+1,5) edge node {$\vec{v}^{s}_{1}$} (6+1,6);
\path [->,thin,opacity=1,color=black,left] (4.5+1,3) edge node {$\vec{v}^{s}_{2}$} (4+1,2);
\shade[ball color = blue,opacity=1] (2+1,4) circle (0.4);
\shade[ball color = blue,opacity=1] (5.5+1,5) circle (0.4);
\shade[ball color = blue,opacity=1] (4.5+1,3) circle (0.4);
\path [-,thin,opacity=1,color=black] (4.5+1,4) edge [out=90, in=160]  (5.25+1,4.5);
\node [color=black,left] at (4.8+1,4.7) {{$\theta$}};
\node [color=white] at (2+1,4)    {{Z$^f_1$}};
\node [color=white] at (5.5+1,5.) {{Z$^s_1$}};
\node [color=white] at (4.5+1,3)  {{Z$^s_2$}};
\end{tikzpicture}
\caption{Definition of the relevant kinematic observables in the rest frame of the intermediate system Z$^f_{2}$.\label{fig:obs}}
\end{figure}
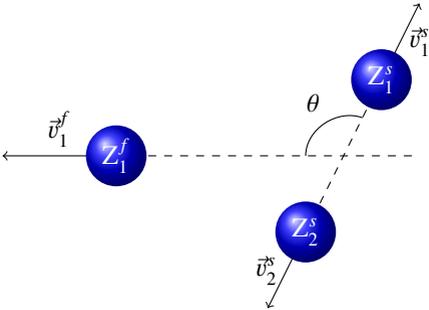

Once the sequence of splittings is known event by event, fragments can be sorted according to their order of production and
the intermediate system can be reconstructed.
Let us now call Z$^f_{1}$ and Z$^f_{2}$, the two nuclei coming from the first splitting. 
The heaviest fragment Z$^f_{2}$ breaks later in Z$^s_{1}$ and Z$^s_{2}$. 
Mean charges of all fragments are presented in Tab.\ref{tab:charges}. For each reconstructed splitting, we also compute
the charge asymmetry $\langle\delta Z^i\rangle = \langle(Z^i_2-Z^i_1)/(Z^i_2+Z^i_1)\rangle$.
Mean charges and asymmetries are comparable for all beam energies (Tab.\ref{tab:charges}).
In addition, the mean asymmetry of the first splitting $\langle\delta Z^f\rangle$ is significantly larger than the quasi-symmetric entrance channel.
It is a strong indication that the first stage of the reactions is an incomplete fusion of projectile and target nuclei,
leading to the formation of heavy composite systems with atomic numbers at least as large as the values of $\langle Z_{src}\rangle$ 
(no attempt was made to correct fragment charges for pre- or post-scission evaporation of charged particles).

\subsection{Decrease of the inter-splitting time}

To measure the inter-splitting time, we used the correlation between the inter-splitting angle $\theta$ 
and the relative velocity of the second splitting: $v^{s}_{12} = \parallel \vec{v}^{s}_1 - \vec{v}^{s}_2 \parallel$ (Fig.\ref{fig:obs}).
For long inter-splitting times the second splitting occurs far from the first emitted fragment.
The relative velocity $v^{s}_{12}$ is then only determined by the mutual repulsion between
Z$^s_{1}$ and Z$^s_{2}$ and should not depend on the relative orientation of the two splittings.
In other words, for long inter-splitting times $v^{s}_{12}$ should be independent of $\theta$.
However, for short inter-splitting time the second splitting occurs close to the first emitted fragment.
The relative velocity $v^{s}_{12}$ is modified by the Coulomb field of Z$^f_{1}$ and depends on the relative 
orientation of the two splittings. In this case, $v^{s}_{12}$ should present a maximum for $\theta=90$\textdegree.
We used this Coulomb proximity effect as a chronometer to measure the inter-splitting time $\delta t$.

\begin{figure}
\begin{minipage}[t]{.45\linewidth}
\center\includegraphics[width=1\linewidth]{\imgpath/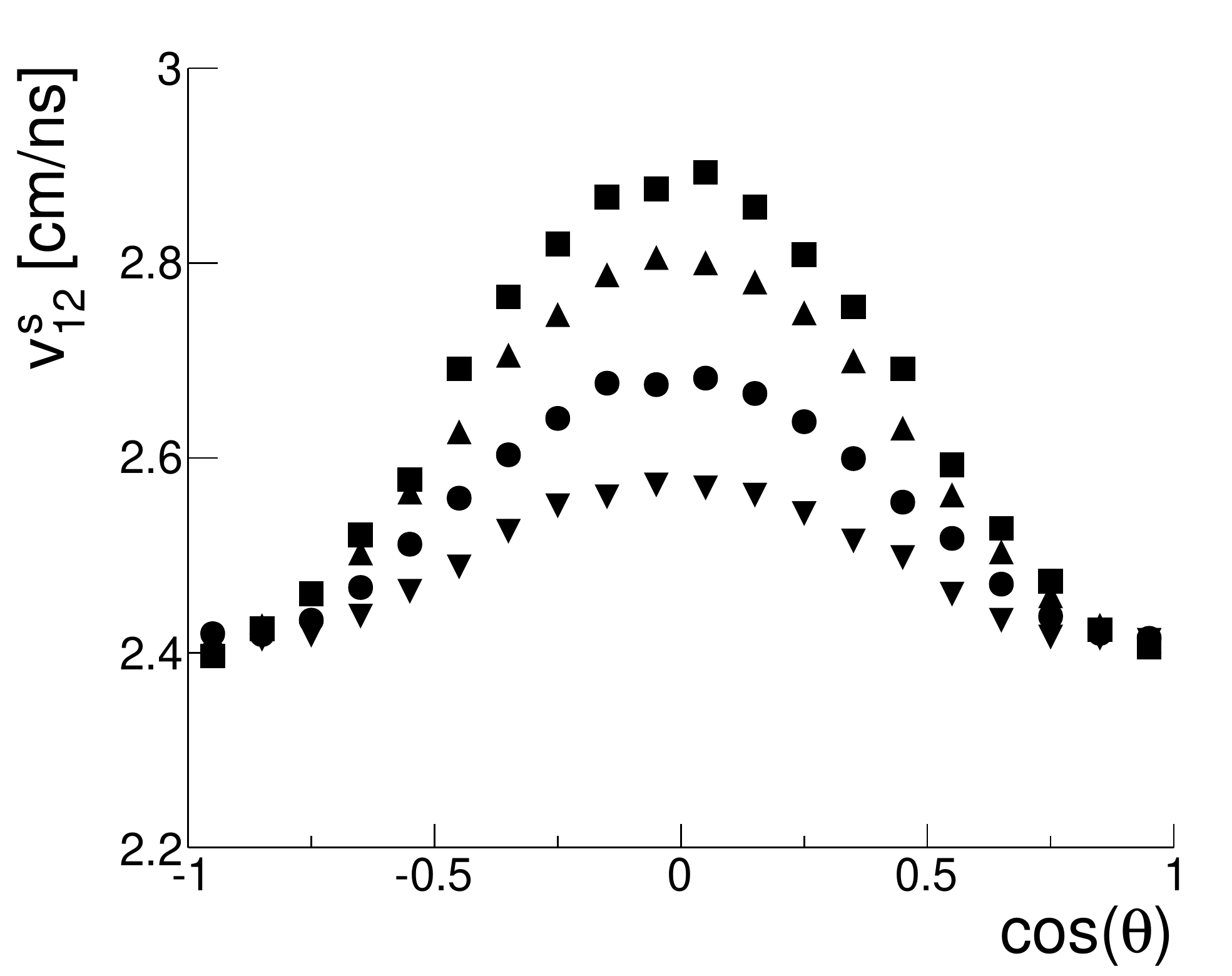}
\caption{Correlation between the inter-splitting angle and the relative velocity of the second splitting for: 
(triangles down) 12\mev, (circles) 15\mev, (triangles up) 18\mev, (squares) 20\mev.\label{fig:vt}}
\end{minipage} \hfill
\begin{minipage}[t]{.45\linewidth}
\center\includegraphics[width=1\linewidth]{\imgpath/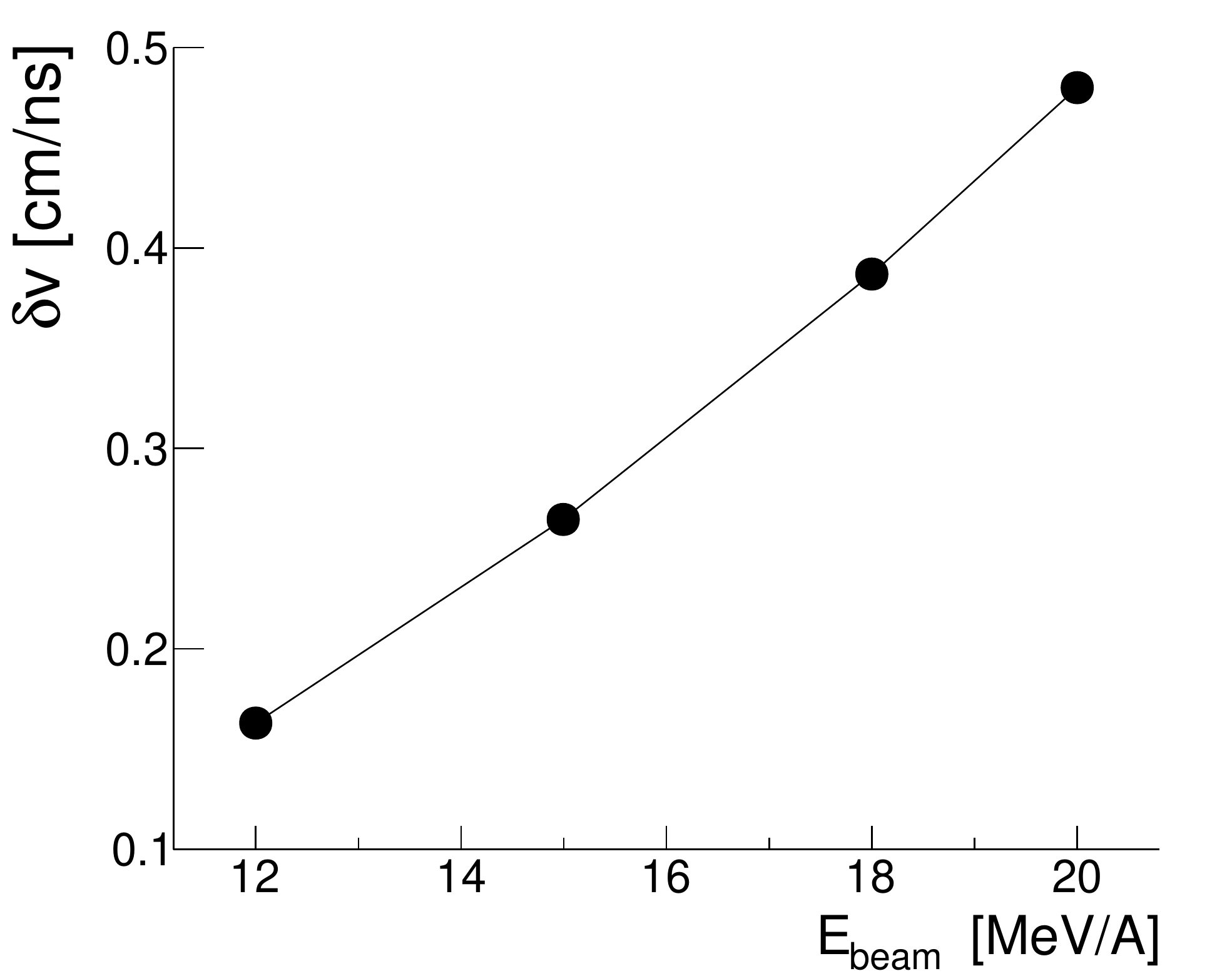}
\caption{Evolution of the Coulomb distortion parameter $\delta v$ as a function of the beam energy for \Xesn central collisions.\label{fig:dv}}
\end{minipage}
\end{figure}

The experimental correlation between $v^{s}_{12}$ and $\theta$ is presented in Fig.\ref{fig:vt}, for all beam energies.
These correlations present a maximum at $\theta\sim90$\textdegree,~which increases with increasing beam energy.
We quantify this effect by the Coulomb distortion parameter $\delta v = v^{s}_{12}(90^\circ) - v^{s}_{12}(0^\circ)$,
which increases with the beam energy (Fig.\ref{fig:dv}). It indicates that the second splitting occured closer and closer to the first emitted fragment.

\begin{figure}[ht]
\sidecaption
\includegraphics[width=0.5\linewidth]{\imgpath/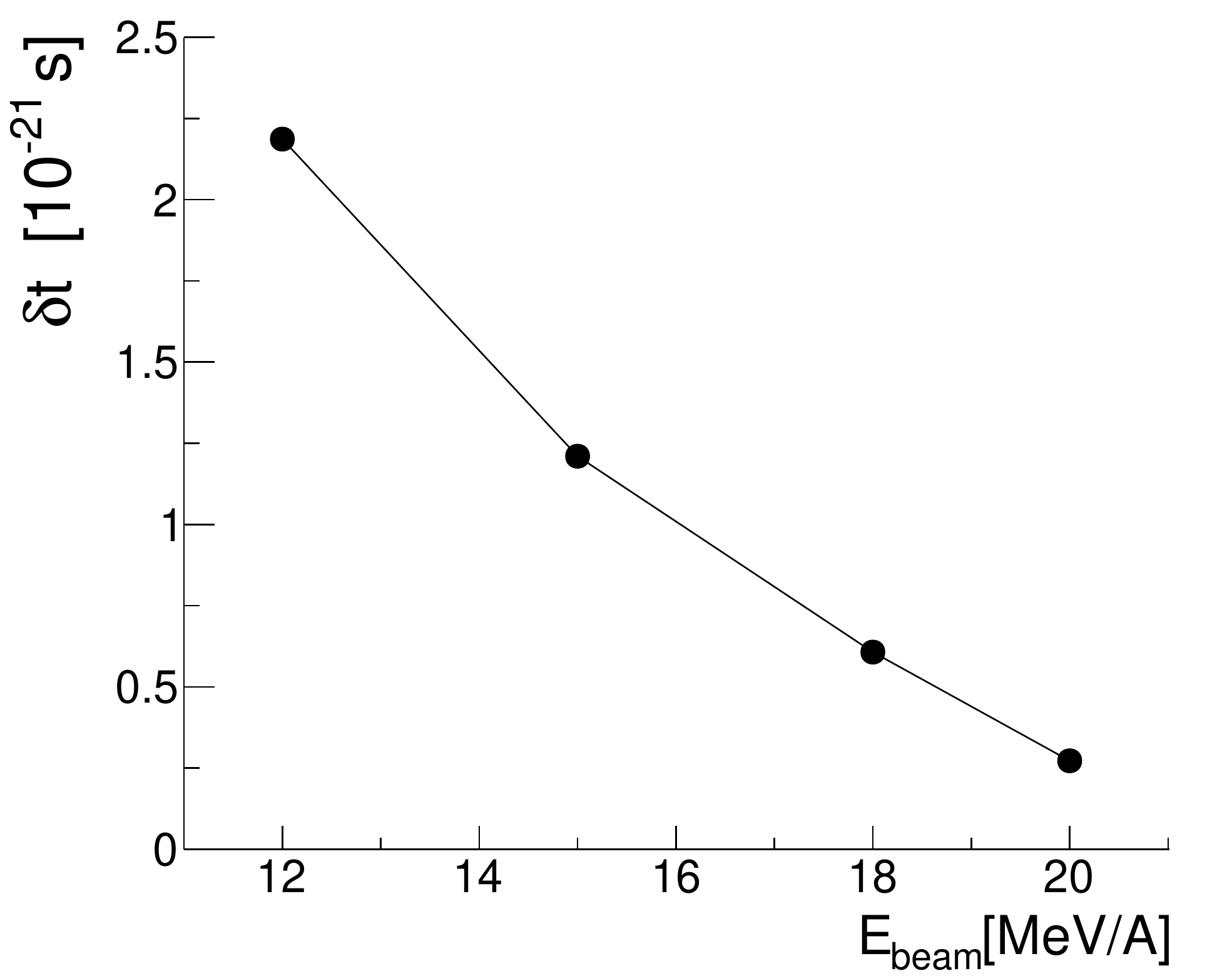}
\caption{Evolution of the inter-splitting time $\delta t$ as a function of the beam energy for \Xesn central collisions.\label{fig:dte}}
\end{figure}

To translate $\delta v$ in terms of inter-splitting time $\delta t$, we performed simple Coulomb trajectory calculations
for three fragments using mean charges given in Tab.\ref{tab:charges}.
We simulated sequential break-ups and we computed $\delta v$ by varying $\delta t$ to get a calibration function.
Finally, we obtained the evolution of the inter-splitting time as a function of the beam energy (Fig.\ref{fig:dte}).
At 12 \mev, \Dt is of the order of $2\times10^{-21}$ s and decreases by a factor eight over the studied bombarding energy range.
Our trajectory calculations show that below $\delta t \sim 0.5\times10^{-21}$ s it is no longer meaningful to speak of sequential fission.
Indeed, the two nuclei resulting from the first splitting do not have sufficient time to move apart beyond the range of the nuclear forces 
before the second splitting occurs. This inter-splitting time is reached around 18 \mev.
Therefore, the decrease of \Dt with increasing beam energy shows the continuous evolution of the decay mechanism,
from hot sequential fission toward multifragmentation.

\section{Conclusion}
In this contribution, we have investigated the three-fragment exit channel in \Xesn central collisions from 12 to 20 \mev.
These fragments arise mainly from two successive splittings which are compatible with sequential fissions of heavy composite systems.
We estimated the time between the two successive fissions by Coulomb chronometry.
Starting from $\delta t \sim 2\times10^{-21}$s at 12 \mev, the inter-splitting time decreases by a factor 
eight over the studied bombarding energy range, becoming compatible with simultaneous multifragmentation above 18 \mev.

\begin{acknowledgement}
The authors would like to thank the staff of the GANIL Accelerator
facility for their continued support during the experiments. D. G.
gratefully acknowledges the financial support of the Commissariat
\`a l'\'energie Atomique and the Conseil R\'egional de Basse-Normandie.
The work was partially sponsored by the French-Polish agreements IN2P3-COPIN
(Project No.~09-136). 
\end{acknowledgement}

\bibliography{articles}

\end{document}